# White-box Inference Attacks against Centralized Machine Learning and Federated Learning

Jingyi Ge

College of Information Science and Technology, Donghua University,201620,Shanghai, China

2222138@mail.dhu.edu.cn

**Abstract.**

With the development of information science and technology, various industries have generated massive amounts of data, and machine learning is widely used in the analysis of big data. However, if the privacy of machine learning applications' customers cannot be guaranteed, it will cause security threats and losses to users' personal privacy information and service providers. Therefore, the issue of privacy protection of machine learning has received wide attention. For centralized machine learning models, we evaluate the impact of different neural network layers, gradient, gradient norm, and fine-tuned models on member inference attack performance with prior knowledge; For the federated learning model, we discuss the location of the attacker in the target model and its attack mode. The results show that the centralized machine learning model shows more serious member information leakage in all aspects, and the accuracy of the attacker in the central parameter server is significantly higher than the local Inference attacks as participants.

**Key words:** machine learning, federated learning, white-box inference attacks, stochastic gradient descent.

## 1 Introduction

### 1.1 Introduction

Machine learning is the technical core of the vigorous development of contemporary artificial intelligence and the basic way of computer intelligence.In machine learning, users can get output results with arbitrary input through general algorithms. Machine learning will collect and learn previous data in this way, and improve the algorithm of the computer itself, so as to effectively optimize the performance of computer programs.

Machine learning is widely used in the analysis and processing of big data because of its superior information processing ability, such as medical diagnosis, weather prediction, economic research, mine engineering, and so on.

However, the model training calculation process, the activation function and so on. Training data privacy: In machine learning, the sample data includes some personalized identifiable information (Personality Identifiable Information, PII), which can represent user attributes, such as email address, address, surname and other user identity attributes. Privacy (3) predict conclusion: machine model will directly calculate the prediction of users, the predicted information may be input the privacy information, such as medical diagnosis model can predict the probability of patients with a disease, the predicted personal information may be used for malicious diagnosis service providers.If the privacy rights in machine learning applications cannot be guaranteed and pose a threat to the security of users' personal privacy information, it is not only for the users using the service, but also for the service providers.

**1.2 Research status quo**

With the expansion of the application of information science and technology, the research has also made vigorous progress in depth and breadth, and the research of privacy-related issues has produced numerous achievements and insights in machine learning and deep learning algorithms.

**Attack strategy.** Shokri[2]These are the first researchers to propose member reasoning attacks. When the black box frame of the target for the internal logic function and operation process, they let the attacker trained on the shadow models , shadow model has similar distribution and architecture with the target, so its behavior on the training data and the target model on the training data behavior more or less similar, can effectively achieve the attack effect. The output statistical characteristics (such as entropy) can be used to perform member inference. In the face of limited access and insufficient samples in black-box inference attacks, we use the fast attack for high-precision member inference, Peng[4]et al launched a new member inference attack method, based on principal component (PCA) member reasoning attack (PCA-based attack), the low migration of the previous experimental model is effectively improved, and this member reasoning attack can be member reasoning attack without the target model information.

Fredrikson[5] designed the inversion attack (Model Inversion) technology firstly, they used the attacker as the central parameter server of the federated learning model, with the maximum posterior probability and model prediction confidence as the attack mechanism.Thereafter, Fredrikson's[6] model reversal attack is improved. In order to apply their attack to the non-linear target model, and the target is regarded as the parameter input, the loss function is optimized by using the reconstruction attack. The improved reversal attack can effectively handle the discrete data. Hitaj[7]et al's research applied reversal attacks to attackers as federated learning participants, who reconstruct the input samples of participants in the model of the model by simulating data samples overlapping the target distribution using a generative adversarial network (GAN). The restoration of a certain category of pictures is realized on the multilayer neural network. This is an expansion and deepening of the attack proposed by Fredrikson et al.

Ateniese[12] et al made usage of an attack according to property, they launched a known target model intrinsic function calculation process and training mode under the premise of inferred machine learning participants sample data and some original model data of the specific properties of the correlation inference attack, and named the attack attribute (Property) inference attack, and in the speech data set inferred accent information identification task reflects the effectiveness of the attack.Their attribute inference attack is captured by Ganju[13]et al realized the expanded application in the fully connected network. Blanchard[15] et al set the malicious participants[16] as the attacker, with the central parameter server using the linear aggregation method as the attack target, found that the target transmission of messages fabricated according to the local model, or even arbitrary data and target transmission, can achieve the training intervention of the model.Experiments show that when a large number of participants in the distributed learning model hide or malicious participants who are prone to leak information, even less capable attackers can hijack the target model and implement Byzantine attacks.

**Defense strategy.** Dwork[19]et al first put forward the concept and technology of differential privacy defense technology. They used noise to intervene the output of the model, which blurred the attack effect of attackers targeting the model output, making it impossible for attackers to identify multiple target data sets, and making it difficult to make a correct judgment on the membership of the data points.Shokri[20] used the stochastic gradient descent (SGD) optimization method, the differential privacy protection technology in the process of distributed machine learning, but their privacy protection method need to randomly selected on the gradient greater than set threshold met $\varepsilon$- differential privacy Laplace noise, and strict privacy budget controls each privacy consumption until exhausted, shut off data access rights, this method is difficult for practice.

In summary, Machine learning has hidden privacy information security risks in terms of model construction and input application, Many scholars' research on federated learning, distributed learning and other environments has accelerated the rapid development of machine learning technology, We use a white-box inference attack model as a means of comparing the training characteristics of centralized machine learning models and federated learning models, And to explore the corresponding risks generated, To refine the study of machine learning privacy risks, To supplement the offensive and defensive strategies in the field of centralized machine learning, Finally, the infinite possibilities for the future applications of machine learning models in various scenarios are explored.

**1.3 Study content and chapter arrangement**

Come in for Nasr[25] et al's study for the privacy risk of deep learning, this paper used a white box member reasoning attack simulation experiment, aimed to explore in the machine deep learning environment, centralized machine learning and federal learning setting, for the level of information leakage and affect the level of the model, and will compare the two learning mechanism characteristics. This paper will be respectively for centralized machine learning and federated learning model member inference attack, using stochastic gradient descent optimizer, model gradient vector as an important attack index, the final simulation results can show the white box member inference attack model for the two target model membership inference accuracy, namely the current learning model in what leaked their own training data.Our evaluation directions include: supervised attacks and unsupervised attacks, inference attacks on the model and its newer versions, passive member reasoning attacks, and active member inference attacks. The experimental results evaluate the model performance based on the member inference attack accuracy score and the True / False positive (ROC) curve.

The first part of this paper introduces the privacy rights in machine learning and the existing attack and defense strategies for machine learning. The second part of the related knowledge, the list of the discussed problems and simulation experiments of the background knowledge and the function basis and its definition. The third part, the model and algorithm, shows the workflow and fundamentals of the white-box member inference attack model used in this paper.The fourth part of white box inference attack performance evaluation introduces the experimental related work setting and process, model attack performance evaluation index and attack simulation results for centralized machine learning model and federated learning model.

**2 Related knowledge**

## 2.1 Member Reasoning Attack

The degree of privacy information leakage of a certain model can be defined as: the attacking party can get one or some private data through this mode. The former shows an increased utility, while the latter reflects a privacy loss. This paper uses white-box member inference attacks to quantify this privacy leakage.

Generally speaking, the purpose of an algorithm for member inference attacks is to reason about the identity of a particular data (a member instance or a non-member instance) in the target training set.In practice, attackers with different training premises use member inference attacks to infer the membership of a given data to the target model dataset.Part of the data belonging to the target model data set is observed and used by the attackers to infer more relevant information of the target data set. Therefore, under the attack mode of member reasoning, the important private information of the training data is likely to be displayed through the leakage degree of the target model.

This paper uses the way of member reasoning attack to obtain more visual and valuable machine learning model mechanism and vulnerabilities, the results can reflect the information leakage degree and privacy security of the machine model in the learning process.

## 2.2 Shadow training techniques and white-box reasoning attacks

**Shadow training technique (Shadow Training Technique).** When the attacker cannot obtain the internal algorithm, the attack feature data cannot be obtained, so the attacker can only start with the output constantly updated by the target model of the neural network layer at any input.Shokri[2]In order to effectively conduct member reasoning attacks, et al. used shadow training techniques to deal with such situations.

**White-box membership inference attack.** In this environment, the attacker can observe not only the model output f (x; W), but also the period operation and all parameters involved in the training process, including all hidden layer output hi (x), so the attacker with white box permission can effectively expand the attack environment of the attacker with black box permission.

## 2.3 Supervised attacks and unsupervised attacks

Supervised and unsupervised learning depends on whether an attacker has prior knowledge such as a part of the target dataset or a sample codistributed with the target sample.When have this knowledge, we will use the way of supervision let the attack model on the known data to build the attack model, that is to let the reasoning model directly learning attack data points and target model members of the membership, in such supervised learning mode to build the reasoning attack became a supervised attack.

However, when the attacker does not have knowledge and pre-training conditions on the internal structure and sample distribution of the target model, we choose to build an unsupervised attack model, predict more information about the target data set according to the underlying output of the target model, and develop member inference attacks.

### 2.4 Stochastic Gradient Descent (SGD) optimizer

Stochastic gradient descent (SGD) algorithm is one of the deep learning optimizers, and it is also a relatively basic neural network optimization method.The algorithm repeatedly update model parameters W as the gradient drop, by calculating the loss function value gradient, iterative weight and bias, reduce the empirical expectation on the training set D, make it tend to zero, so that the model parameters constantly approaching the expression of the real data, as shown in formula (1), for stochastic gradient descent (SGD) algorithm working principle, which is the classification model f loss function.

$$\min_{W} E(x, y) \sim D[L(f(x; W), y)] \tag{1}$$

Stochastic gradient descent algorithm will leave marks for the parameter loss function gradient of each trained sample, which is the basis of reasoning attack, the white box inference attack model and use of stochastic gradient descent optimization algorithm, these sample markers can make the model gradient vector of all parameters on the attack of the target easy to be observed, which becomes our important attack index.

### 2.5 Passive attack and active attack

Passive or active white-box member inference attack mode depends on whether the attacker passively receives the update parameter gradient of the observation model or actively influences the training process of the target model.

**Active attack (global attacker and local attacker).** In the active attack mode, the attacker will participate in the target model training process, and obtain the corresponding training set member information through the active influence of their training parameters, based on which the reasoning attack is implemented.Due to the structural characteristics of federated learning, active attacks are often applied to them. The central parameter server will distribute the parameters before the start of each training, collect the local model parameters uploaded by each participant and aggregate and update the global models, and each training stage can contribute to the attacker's attack.

$$W \leftarrow W + \gamma \frac{\partial L_x}{\partial W} \qquad (2)$$

## 2.6 Centralized learning

In this environment setting, all data are trained in the central parameter server set, which includes public general data as well as some private data. The attack model is able to observe the complete independent learning model and the training results of each output.

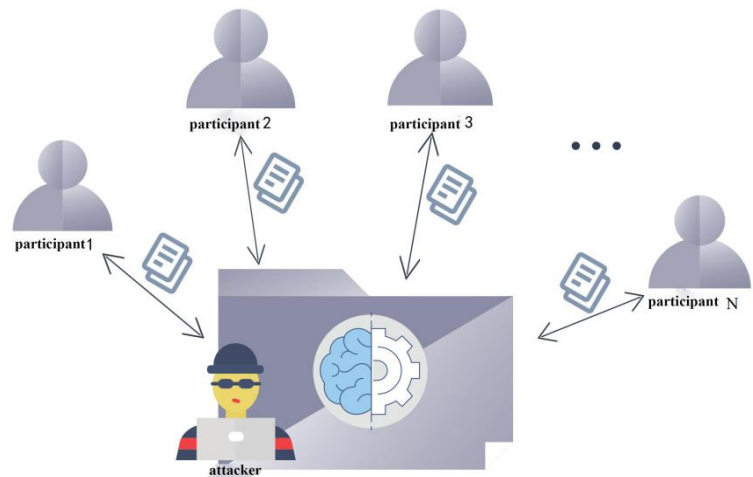

**Fig 1** Centralized machine learning.

In the experimental study of this paper, training with the new dataset d to get model updates, These fine-tuning (Fine-tuning) models are often caused by the effects of those private data, In the following training sessions, The attacker can observe the fine-tuning of the new independent model f and its training results, Member inference for the new dataset d, besides, The attacker will also make inference attacks on the two versions of the data set before and after the fine update. To get more member information on your private data, Restore the important privacy involved.

## 2.7 Federated Learning

In order to achieve a certain accuracy and have stronger reference value, machine learning needs to cover enough large dimension training samples, but in practical application, data collection not only need high density sample transmission, also exacerbated the privacy information crisis, because the larger the base data often contains more important information in the field of information, this undoubtedly enhances the privacy risk. Under the federated learning framework, the central parameter server each participant in federated learning training has their own training set, they download global parameters in the beginning, and training in the local model, they do local update and upload back to the central parameter server, parameter server processing N participants of the update average, data aggregation sharing save the latest parameter version of the global model. Participants' sharing content is limited to parameters rather than specific data.

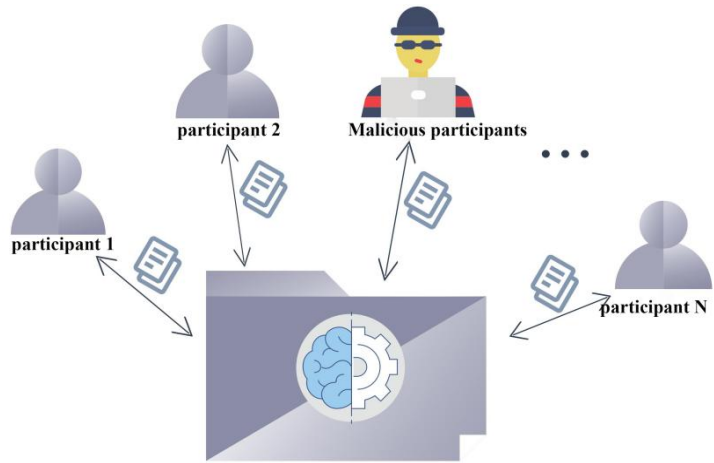

**Fig 2** Privacy risks of centralized machine learning.

## 3 Models and algorithms used

### 3.1 White-box membership inference attack model

For the two types of attackers classified according to having or without background knowledge, we trained the attack model to perform the attack in different ways.We performed attack simulation experiments using the Alexnet model trained on the CIFAR-100 dataset as input of the attack model, and the following white-box member inference attack model.

## 3.2 Principles and algorithms

In the white-box inference attack model we use, the tag components are built on a fully connected network. Aggregate all local weights received from the previous layer, form inputs to nonlinear functions, form the output values of each module of the convolutional layer, and optimize model convergence.The white box reasoning attack model recombines the output of each submodule component, combines the output of all feature extraction components through the encoder components, the encoder output constitutes the attack output of the model, the output is divided into "members" and "non-members", we put the accuracy of the model for the unknown stronghold (predicted members, non-members) as our basis to judge the attack performance of the model.

## 4 Experiments of the white-box membership inference attack

### 4.1 Experimental setup

The equipment prepared for the experiment is the Intel core i7-8650U CPU memory 16.0GB of the computer, the experimental language Python.The experiment uses Pytorch to complete the neural network, which is easy to define, which helps us to easily and quickly establish relatively small projects.

**Attack Model.** ReLU activation function is defined as formula (5), where yi is the model output as a non-saturated activation function, its unilateral inhibition ability to the output can solve a certain "gradient disappearance" problem, and effectively improve the model convergence speed to help effectively realize non-linear activation transformation. Otherwise, ReLU activation function can sparse the model well, and such sparsity facilitates experimental fitting to the data.

$$f(y_i) = \begin{cases} y_i, & y_i > 0 \\ 0, & y_i \leq 0 \end{cases} \qquad (6)$$

**Performance evaluation indicators.** For the white-box model used in this paper, we used two —— centralized machine learning and federated model —— with centralized performance evaluation indicators to form a more comprehensive evaluation system, to help improve the learning characteristics of the two learning models and the differentiation of member information leakage level.

Table 1 Parameter Control Table.

| parameter | Parameter name |
| --- | --- |
| f | object model |
| h | Attack model |
| W | Target model parameters |
| D, D' | Member and non-member datasets |
| Pr | Precision--Recall Rate |
| γ | Adverse update rate |

## 4.2 Analysis of the simulation experiment results

Deep learning has two main training algorithms. In this paper, we first show the simulation results analysis of the centralized machine learning, and then provide the simulation results and analysis of the federated learning objective model.

**Simulation results analysis of white-box membership inference attacks for centralized machine learning.** For the attack of centralized machine learning model, this paper will evaluate the white box attack model from the following dimensions: based on the attack model's understanding of target training mode, sample distribution, attack and defense mechanism, starting from the two presets of supervised attack and unsupervised attack.In supervised attacks, the output of different layers from the trained attacker attack model and the gradient of the classification model; in unsupervised attacks, we use shadow training techniques; and in a special case, we update multiple versions of the target model simultaneously.

*Supervised attacks.* In the context of supervised attacks, we train the target and attack models using a pre-trained CIFAR100 dataset. Therefore, the white-box inference attack model understands the subset and sample distribution of the target, and can start the member reasoning attack from the gradient and number of layers of the target model.

Table 2 Shows the member inference attack accuracy for the outputs of different layers of the CIFAR1000-Alexnet centralized learning model

| output layer | Member prediction accuracy |
| --- | --- |
| The bottom third layer | 72.88% |
| The penultimate layer | 74.06% |
| The final layer | 74.61% |

When the training of the target model enters the last few layers, it will contain a lot of training information from its previous layers, which makes the more data information the model stores and covers more parameters, which is one of the reasons why the output of the final layer of the target model leaks more membership information.

Since the deep neural network contains large-scale parameter data beyond the dimensions that the target model is correctly generalized, the parameter gradient in the target model will show differences that attackers can easily distinguish between.

We used the CIFAR00 dataset to train the Densenet model with model parameter scale 25.62M and parameter scale 1.7M two models as the target model.

The experiment also performs distribution statistics for the gradient norm of membership of the output of each output class of the target CIFAR100-Alexnet centralized machine learning model.

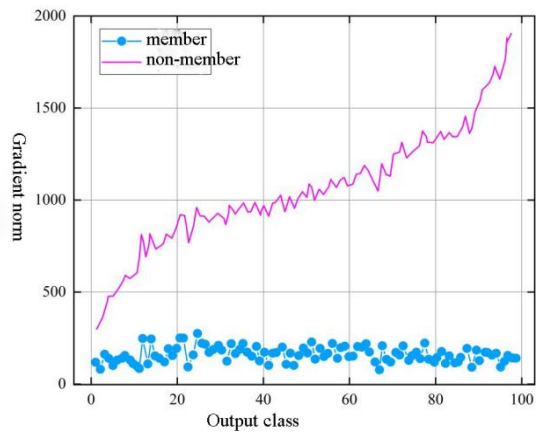

**Fig5** Gradient norm distribution of output members and non-members of each target CIFAR100-Alexnet centralized machine learning model[25].

Since the gradient distribution of members of CIFAR100-Densenet and non-members of CIFAR100-Densenet model is more different than that of CIFAR100-Resnet model, the attack accuracy on Densenet architecture is better both from model output and model gradient perspective.

**Table 3** Attack models attack the three target models with different architectures trained on the CIFAR100 dataset, respectively for the output of the model and the member inference accuracy of the gradient.

| object model | | Attack accuracy | |
| --- | --- | --- | --- |
| Dataset | Model architecture | target outputs | target gradients |
| CIFAR100 | Alexnet | 74.61% | 75.09% |
| CIFAR100 | Resnet | 62.20% | 64.34% |
| CIFAR100 | Densenet | 67.72% | 74.31% |

*Unsupervised attacks.* In order to assign members of the target model to non-member examples, according to the gradient model value, we give the spectral clustering algorithm with "member" cluster sample, and the other samples we preset as "non-member" to facilitate the distribution of members and non-member samples to determine the member inference accuracy of unsupervised attacks.

*Fine-tuned models.* Such attack observation was chosen because we want to expand the attack model from one to multiple to improve its performance. And in fact, model fine-tuning is often influenced by important private data.

**Table 4** Attack Accuracy of attacking fine-tuned models trained on the CIFAR100 dataset in centralized machine learning.

| Dataset | Model architecture | Non-member in Dataset D | Non-member in Dataset DΔ |
| --- | --- | --- | --- |
| CIFAR100 | Alexnet | 75.38% | 71.36% |
| CIFAR100 | DenseNet | 74.61% | 71.50% |

*Passive attack.* The experiment will make the Alexnet model trained on the CIFAR100 dataset as the most target model, and set the position of the attacker to the central parameter

server.First, our attack starts with the training phase of the model.Attacks follow five discontinuous training sessions against the target model, from 5 to 300.

Table 5 Attack accuracy of passive global attacker's attacking different training stages of the Alexnet federated learning model trained on the CIFAR100 dataset.

| Training stage | Attack accuracy |
| --- | --- |
| 5 10 15 20 25 | 57.32% |
| 10 20 30 40 50 | 76.47% |
| 50 100 150 200 | 79.50% |
| 100 150 200 250 300 | 84.89% |

*Active attack.* Due to the federated learning mode, the central parameter server for each update data aggregation will negatively affect the reasoning accuracy of attack model, so in order to get more training set data information of target participants, we use the same attack model, preset global attacker active isolation target training participants, intervention in his training process, to hinder its data upload and receive.

Table 6 The attack accuracy of passive and active white-box members for target models (four participants) trained on the CIFAR100 dataset at different locations under the federated learning architecture.

| Target model | | Global attacker | | Local attacker |
| --- | --- | --- | --- | --- |
| Dataset | Model architecture | Passive | Active | Passive |
| CIFAR100 | Alexnet | 84.98% | 88.52% | 72.88% |
| CIFAR100 | Densenet | 77.43% | 82.90% | 71.98% |

Active global attacker in the implementation of the attack, will actively block part of the central parameter server issued parameters, which makes the target participants not only cannot share with the central parameter server, also and other participants smooth aggregation, which strengthens the attack model gradient superposition of the local SDG algorithm, causing the target local model internal members and non-member example become easier to identify.

## 5 Conclusion

In contrast to the leakage of training and learning information and related privacy vulnerabilities in the training and learning characteristics of the internal member samples of centralized machine learning and federated learning models, we use different pre-trained attack models to distinguish their access to and knowledge of the internal function, training patterns, learning methods, distribution, and training set information of the target model.

Experimental data show we use the white box members of the inference attack evaluation results: under the premise of using such attack model for attack, training late centralized machine learning model in all aspects show greater members of information leakage, neural network layer, gradient two attack indicators showed similar attack results, and for the latter, as a global model of passive attacker members of the reasoning accuracy will be significantly higher than the participants of the local attack architecture.

Federal machine learning broke the traditional centralized machine learning centralized data control deadlock, to personal local data privacy security issues provides new ideas, through the data communication and sharing bridge, improve the efficiency of the machine learning model, accuracy and high reference value, in the era of all Internet, machine learning technology inspired wide attention and innovation, based on attack and defensive research, spawned multi-dimensional application of attack strategy[28]. And for the powerful attack ability of white-box membership inference attacks, there are now some adversarial defense strategies[28]Still helpless about it means that information security in personally sensitive areas is still an urgent issue for users using related technology products, because it can even lead to serious property losses, which undoubtedly poses great obstacles to the use of machine learning in daily life.

Reasoning in this paper, the field of machine learning white box members against sensitive sectors in order to improve information security and user privacy information security protection problems to provide important insights and discuss, for the big data age of machine learning is widely used in everyday life and prosperity and development of solid foundation, to speed up the steps of 5 g era, it can better popularize machine learning technology to all aspects of social life, so that the general public can enjoy the convenience and have less worries.In addition, more loopholes and learning features of machine learning are waiting for us to explore and further study.

# 6 References

[1] SHOKRI R, STRONATI M. SONG C, et al. Membership inference attacks against machine learning models[C]//IEEE Symposium on Security and Privacy.2017:3-18.

[2] Salem A, Zhang Y, Humbert M, et al.Ml-leaks:Model and data independent membership inference attacks and defenses on machine learning models[J].arXiv preprint arXiv:1806.01246,2018.

[3] FREDRIKSON M, LANTZ E, JHA S, et al. Privacy in pharmacogenetics: An end-to-end case study of personalized warfarin dosing[C]//USENIX Security.2014:17-32.

[4] FREDRIKSON M, JHA S, RISTENPART T. Model inversion attacks that exploit confidence information and basic countermeasures[C]//Proceedings of ACM SIGSAC. 2015:1322-1333.

[5] ZHU L, HAN S. Deep leakage from gradients[C]//Advances in Neural Information Processing Systems. 2019:17-31.

[6] JUUTI M, SZYLLER S, MARCHAL S, et al. PRADA: protecting against DNN model stealing attacks[C]//IEEE European Symposium on Security and Privacy.2019:512-527.

[7] ATENIESE G, MANCINI L V, SPOGNARDI A, et al. Hacking smart machines with smarter ones: How to extract meaningful data from machine learning classifiers[J]. International Journal of Security and Networks, 2015, 10(3): 137-150.

[8] ATENIESE G, MANCINI L V, SPOGNARDI A, et al. Hacking smart machines with smarter ones: How to extract meaningful data from machine learning classifiers[J]. International Journal of Security and Networks, 2015, 10(3): 137-150.

[9] MELIS L, SONG C, DE CRISTOFARO E, et al. Exploiting unintended feature leakage in collaborative learning[C]//IEEE Symposium on Security and Privacy. 2019:691-706.

[10] BLANCHARD P, EL MHAMDI E M, GUERRAOUI R, et al. Machine learning with adversaries: Byzantine tolerant gradient descent[C]//Neural Information Processing Systems.2017:118-128.

[2] BONAWITZ K, EICHNER H, GRIESKAMP W, et al. Towards federated learning at scale: System design[C]//IEEE CVPR Workshop.2019:3310-3319.

[11] FUNG C, YOON C J M, BESCHASTNIKH I. The limitations of federated learning in sybil settings[C]//International Symposium on Research in Attacks. 2020:301-316.

[12] DWORK C, ROTH A. The algorithmic foundations of differential privacy[J]. Foundations and Trends in Theoretical Computer Science, 2014, 9(3/4):211-407.

[13] GEYER R C, KLEIN T, NABI M. Differentially private federated learning: A client level perspective[J]. arXiv preprint, 2017:1712.07557.

[14] BHOWMICK A, DUCHI J, FREUDIGER J, et al. Protection against reconstruction and its applications in private federated learning[J]. arXiv preprint, 2018:1812.00984.

[15] NASR M, SHOKRI R, HOUMANSADR A. Comprehensive privacy analysis of deep learning: Passive and active white-box inference attacks against centralized and federated learning[C] // 2019 IEEE Symposium on Security and Privacy (SP). IEEE, 2019 : 739 - 753.

[16] SZEGEDY C, ZAREMBA W, SUTSKEVER I, et al. Intriguing properties of neural networks[J]. arXiv preprint arXiv: 1312. 6199, 2013.

[17] Ping Zhao, Hongbo Jiang, John C. S. Lui, Chen Wang, Fanzi Zeng, Fu Xiao, and Zhetao Li. P3-LOC: A Privacy-Preserving Paradigm-Driven Framework for Indoor Localization. IEEE/ACM Transactions on Networking (ToN), vol. 26, no. 6, pp. 2856-2869, 2018.

[18] Hongbo Jiang, Yuanmeng Wang, Ping Zhao, Zhu Xiao. A Utility-Aware General Framework with Quantifiable Privacy Preservation for Destination Prediction in LBSs. IEEE/ACM Transactions on Networking (ToN), vol. 29, no. 5, pp. 2228-2241, 2021.

[19] Ping Zhao, Jiaxin Sun, and Guanglin Zhang. DAML: Practical Secure Protocol for Data Aggregation based on Machine Learning. ACM Transactions on Sensor Networks, vol. 16, no. 4, pp. 1-18, 2020.

[20] Ping Zhao, Wuwu Liu, Guanglin Zhang, Zongpeng Li, and Lin Wang. Preserving Privacy in WiFi Localization with Plausible Dummy Locations. IEEE Transactions on Vehicular Technology, vol. 69, no. 10, pp. 11909-11925, 2020.

[21] Guanglin Zhang, Anqi Zhang, Ping Zhao, and Jiaxin Sun. Lightweight Privacy-Preserving Scheme in WiFi Fingerprint-Based Indoor Localization. IEEE Systems Journal, vol. 14, no. 3, pp. 4638-4647, 2020.

[22] Ping Zhao, Jie Li, Fanzi Zeng, Fu Xiao, Chen Wang, Hongbo Jiang. ILLIA: Enabling k-Anonymity-based Privacy Preserving against Location Injection Attacks in continuous LBS Query. IEEE Internet of Things Journal, vol. 5, no. 2, pp. 1033–1042, 2018.

[23] Ping Zhao, Hongbo Jiang, Jie Li, Fanzi Zeng, Xiao Zhu, Kun Xie, and Guanglin Zhang. Synthesizing Privacy Preserving Traces: Enhancing Plausibility with Social Networks. IEEE/ACM Transactions on Networking (ToN), vol. 27, no. 6, pp. 2391 – 2404, 2019.

[24] Ping Zhao, Jiawei Tao, Guanglin Zhang. Deep Reinforcement Learning-based Joint Optimization of Delay and Privacy in Multiple-User MEC Systems. IEEE Transactions on Cloud Computing, DOI: 10.1109/TCC.2022.3140231, 2022.

[25] Ping Zhao, Hongbo Jiang, Jie Li, Zhu Xiao, Daibo Liu, Ju Ren, Deke Guo. Garbage in, Garbage out: Poisoning Attacks Disguised with Plausible Mobility in Data Aggregation. IEEE Transactions on Network Science and Engineering (TNSE), vol. 8, no. 3, pp. 2679-2693, 2021.

[26] Ping Zhao, Xiaohui Zhao, Daiyu Huang, H. Huang. Privacy-Preserving Scheme against Location Data Poisoning Attacks in Mobile-Edge Computing. IEEE Transactions on Computational Social Systems, vol. 7, no. 3, pp. 818-826, 2020.

[27] Ping Zhao, Chen Wang, and Hongbo Jiang. On the Performance of k -Anonymity against Inference Attacks with Background Information. IEEE Internet of Things Journal, vol. 6, no. 1, pp. 808–819, 2018.

[28] Guanglin Zhang, Sifan Ni, and Ping Zhao. Enhancing Privacy Preservation in Speech Data Publishing. IEEE Internet of Things Journal, vol. 7, no. 8, pp. 7357-7367, 2020.

[29] Guanglin Zhang, Anqi Zhang, Ping Zhao. LocMIA: Membership Inference Attacks against Aggregated Location Data. IEEE Internet of Things Journal, vol. 7, no. 12, pp. 11778-11788, 2020.

[30] Guanglin Zhang, Sifan Ni and Ping Zhao. Learning-based Joint Optimization of Energy-Delay and Privacy in Multiple-User Edge-Cloud Collaboration MEC Systems. IEEE Internet of Things Journal, doi: 10.1109/JIOT.2021.3088607, 2021.

[31] P. Zhao, Z. Cao, J. Jiang and F. Gao, "Practical Private Aggregation in Federated Learning Against Inference Attack," in IEEE Internet of Things Journal, 2022, doi: 10.1109/JIOT.2022.3201231.

[32] Hongbo Jiang, Yu Zhang, Zhu Xiao, Ping Zhao and Arun Iyengar. An Empirical Study of Travel Behavior Using Private Car Trajectory Data. IEEE Transactions on Network Science and Engineering, vol. 8, no. 1, pp. 53-64, 2021.